\documentclass[aip,amsmath,amssymb,preprint]{revtex4-1}
\pdfoutput=1
\usepackage{graphicx}
\usepackage{dcolumn}
\usepackage{bm}
\usepackage{hyperref}
\usepackage[mathlines]{lineno}
\usepackage[utf8]{inputenc}
\usepackage[T1]{fontenc}
\usepackage{mathptmx}
\usepackage{etoolbox}
\makeatletter
\def\@email#1#2{%
 \endgroup
 \patchcmd{\titleblock@produce}
  {\frontmatter@RRAPformat}
  {\frontmatter@RRAPformat{\produce@RRAP{*#1\href{mailto:#2}{#2}}}\frontmatter@RRAPformat}
  {}{}
}%
\makeatother
\begin{document}

\title{All solution grown epitaxial magnonic crystal of thulium iron garnet thin film}

 \author{Rajnandini Sharma}
    \email{rajnandinisharma.rs.mst18@itbhu.ac.in}
    \affiliation{School of Materials Science and Technology, Indian Institute of Technology (Banaras Hindu University), Varanasi-221 005, India}

     \author{Pawan Kumar Ojha}
    \affiliation{School of Materials Science and Technology, Indian Institute of Technology (Banaras Hindu University), Varanasi-221 005, India}
       
    \author{Simran Sahoo}
    \affiliation{School of Materials Science and Technology, Indian Institute of Technology (Banaras Hindu University), Varanasi-221 005, India}

    \author{Rijul Roychowdhury}
    \affiliation{Surface Physics and Material Science Division, Saha Institute of Nuclear Physics Kolkata, 1/AF Bidhannagar, Sector 1, Kolkata 700 064, India}

    \author{Shrawan Kumar Mishra}
    \email{shrawan.mst@iitbhu.ac.in}
    \affiliation{School of Materials Science and Technology, Indian Institute of Technology (Banaras Hindu University), Varanasi-221 005, India}

    \date{\today}
    
\begin{abstract}
Magnonics has shown the immense potential of compatibility with CMOS devices and the ability to be utilized in futuristic quantum computing. Therefore, the magnonic crystals, both metallic and insulating, are under extensive exploration. The presence of high spin-orbit interaction induced by the presence of rare-earth elements in thulium iron garnet (TmIG) increases its potential in magnonic applications. Previously, TmIG thin films were grown using ultra-high vacuum-based techniques. Here, we present a cost-effective solution-based approach that enables the excellent quality interface and surface roughness of the epitaxial TmIG/GGG. The deposited TmIG (12.2 nm) thin film's physical and spin dynamic properties are investigated in detail. The confirmation of the epitaxy using X-ray diffraction in $\phi$-scan geometry along with the X-ray reflectivity and atomic force for the thickness and roughness analysis and topography, respectively. The epitaxial TmIG/GGG have confirmed the perpendicular magnetic anisotropy utilizing the polar-magneto-optic Kerr effect. Analyzing the ferromagnetic resonance study of TmIG/GGG thin films provides the anisotropy constant K$_U$ = 20.6$\times$10$^3$ $\pm$ 0.2$\times$10$^3$ N/m$^2$ and the Gilbert damping parameter $\alpha$ = 0.0216 $\pm$ 0.0028. The experimental findings suggest that the solution-processed TmIG/GGG thin films have the potential to be utilized in device applications.

\end{abstract}
\maketitle
Magnonics is the study of spin waves-based information processing and transmission \cite{flebus2023recent}. Magnons have the potential to be utilized in more dense logic gates, along with the processing and transport of information simultaneously \cite{chumak2017magnonic}. Magnon's superposition ability makes it a potential candidate for its uses as a qubit in quantum computing \cite{andrianov2014magnon}. There are various magnon carrier systems; some are conducting, and others insulating \cite{chumak2017magnonic}. Conducting magnonic crystals are CoFeB\cite{liu2011ferromagnetic,xu2012tuning}, NiFe(permalloy) \cite{obry2013micro,kalarickal2006ferromagnetic}, and Heusler compounds \cite{pirro2014non,sebastian2012low}. Iron garnets are one class of these insulating magnonic crystals \cite{serga2010yig}. Initially, various fundamental understandings of magnon behaviours like magnon-magnon scattering and magnetic resonance have been considered as possible microscopic origins using these ferrimagnetic insulators \cite{sparks1961ferromagnetic,srivastava1999angle}. Recently, the heterostructure yttrium iron garnet (YIG) found its application in spin pumping. Both exchange and dipolar spin waves are higher-order spin waves in the single-crystal YIG thin films \cite{hurben1995theory}. Soon after its discovery, the experimental study confirmed system has the lowest dissipation (lowest linewidth of the ferromagnetic resonance), making it a promising system for various  applications\cite{cherepanov1993saga}. A recent study shows that in pulsed laser deposition (PLD) grown Pt/YIG, the interfacial spin Hall angle ($\theta_{SH}$) is 0.33 \cite{dai2019observation}. But further advanced processing can be achieved by perpendicular magnetic anisotropy (PMA) in the system \cite{garello2014ultrafast}. YIG has a low anisotropy constant K$_U$ = 1$\times 10^{3}$ N/m$^2$ when deposited on Gd$_3$Ga$_5$O$_{12}$ (GGG) substrate\cite{krockenberger2009layer}. High spin-orbit coupling in rare-earth iron garnets has the potential to resolve this \cite{mccloy2013sublattice}. Complete rare-earth series can form the iron garnets \cite{zanjani2020predicting}. The rare-earth elements have their unique magnetic ordering so they contribute to the ferrimagnetic coupling. This contribution causes compensation temperature, which is the lowest magnetization state. Thulium iron garnet (TmIG) is the rare-earth garnet with Curie temperature (T$_C$ $\approx$ 550 K) that has the lowest compensation temperature $\approx$15 K and room temperature moderate saturation magnetization. Recently, the Pt/TmIG heterostructure with PMA shows the magnetic switching and spin magnetoresistance \cite{avci2017current}, and TmIG/Au/TmIG shows the spin valves properties \cite{vilela2020strain}.

Iron garnet thin films are grown using ultra-high vacuum facilities and require extravagant facilities like PLD, off-centred rf-sputtering, and liquid phase epitaxy (LPE). Few studies have produced polycrystalline iron garnets using solution methods like spin-coating \cite{sharma2023magnetic,sharma2023magneticTm,pena2015yig}. However, the epitaxial thin film growth using spin-coating is not reported to date. This article presents a cost-effective, all-solution-based spin-coating method that uses the substrate's crystal structure to reference and grows an epitaxial thin film of TmIG/GGG. The epitaxial TmIG/GGG has been studied using synchrotron Grazing Incident X-ray diffraction (GIXRD). The confirmation of the epitaxy is presented using GIXRD $\phi$-scan. The topography and elemental analysis of TmIG magnonic crystal are studied in detail. The magnetic study of the good interface quality epitaxial TmIG thin film is reported in the present article.

TmIG thin film on single-crystal gadolinium gallium garnet (GGG) substrate of (111) orientation was deposited using all solution-based spin-coating. To prepare a solution,  nitrates of iron (Fe(NO$_3$)$_3$.9H$_2$O, (98$\%$ purity)) and thulium (Tm(NO$_3$)$_3$.5H$_2$O (99.9$\%$ purity)) in 3:2 ratio was amalgam in the 2-methoxyethanol with 400 mM concentration. The solution was stirred and aged for three days to make it uniform and have a gel-like consistency. The substrate's surface quality should be excellent to deposit the thin film. To clean, the substrate (GGG) was ultra-sonicated with de-ionized water, acetone, and 2-propanol for 30 minutes each. Further, the substrate was plasma-cleaned for 10 min at 10 W in an oxygen atmosphere. The uniformly stirred solution was statically spin-coated on the cleaned substrate at 4000 rpm for 30 sec. The excellent interface and film quality are achieved by heating the spin-coated film in three stages. An excessive solvent was initially evaporated at 363 K for 2 hours in the air on the hot plate. Organic solvent decomposition was processed by heating the film further at 623 K for 30 minutes in a muffle furnace (in air). The final phase formation was achieved by annealing the prepared film at 1223 K for 3 hrs in a tubular furnace with an oxygen environment. The crystal structure of GGG and TmIG is analogous; therefore, the growth of the epitaxial TmIG becomes favorable. 

\begin{figure}
    \centering
    \includegraphics[width=1.0\linewidth]{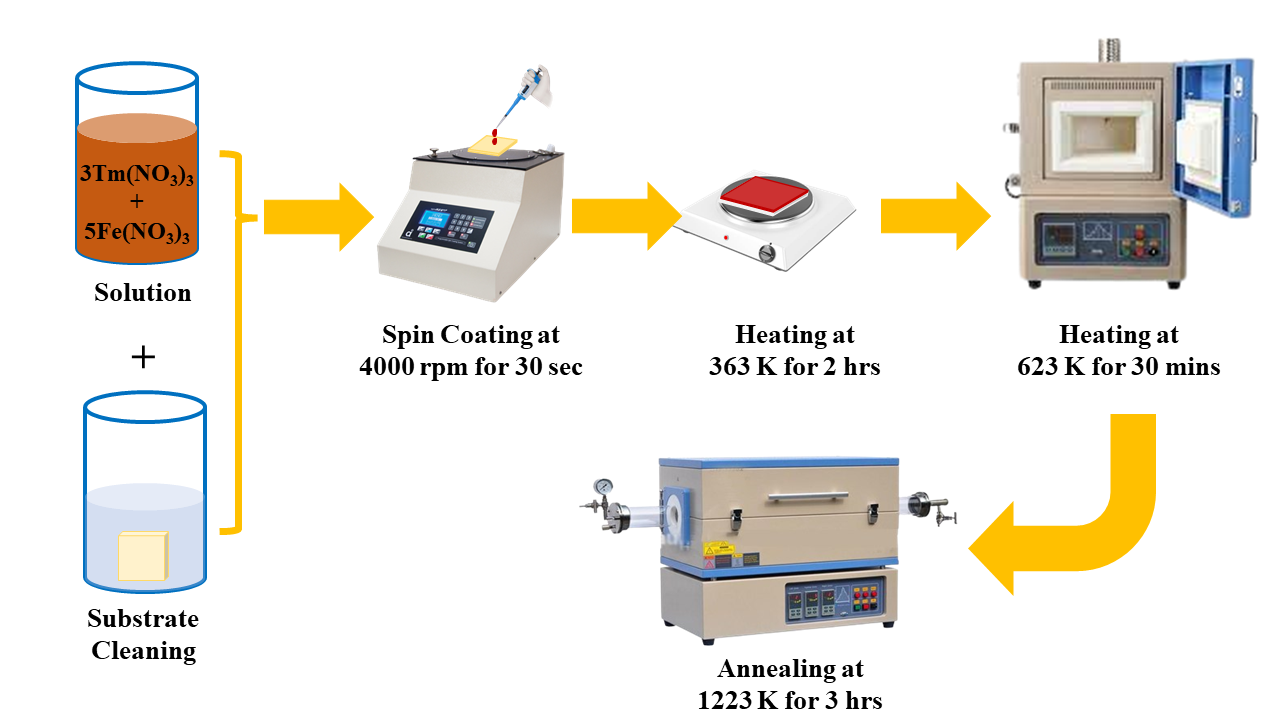}
    \caption{Schematic of the synthesis method of TmIG/GGG thin film.}
    \label{fig:TmIG_GGG_synthesis method}
\end{figure}

The structural confirmation was done using synchrotron grazing-incident x-ray diffraction (GIXRD) using 10 KeV energy of INDUS-2 (BL-13) RRCAT, Indore. Thickness estimation uses X-ray reflectivity (XRR) utilizing Bruker D8 Diffractometer. XRR data was fitted using Parrat32 software utilizing Parratt's formalism \cite{parratt1954surface,feranchuk2003analytical}. The morphology of the thin films was observed using atomic force microscopy (AFM) utilizing a Bruker nano IR microscope. The elemental composition study uses X-ray photoelectron spectroscopy (XPS) using Thermo Fisher Scientific model K alpha using aluminum K-alpha radiation. The magnetic study uses a white LED-based Magneto-optical Kerr effect (MOKE) microscope in polar mode along with room temperature ferromagnetic resonance (FMR), utilizing broadband FMR of Quantum design Phase FMR.

\begin{figure}
    \centering
    \includegraphics[width=1.0\linewidth]{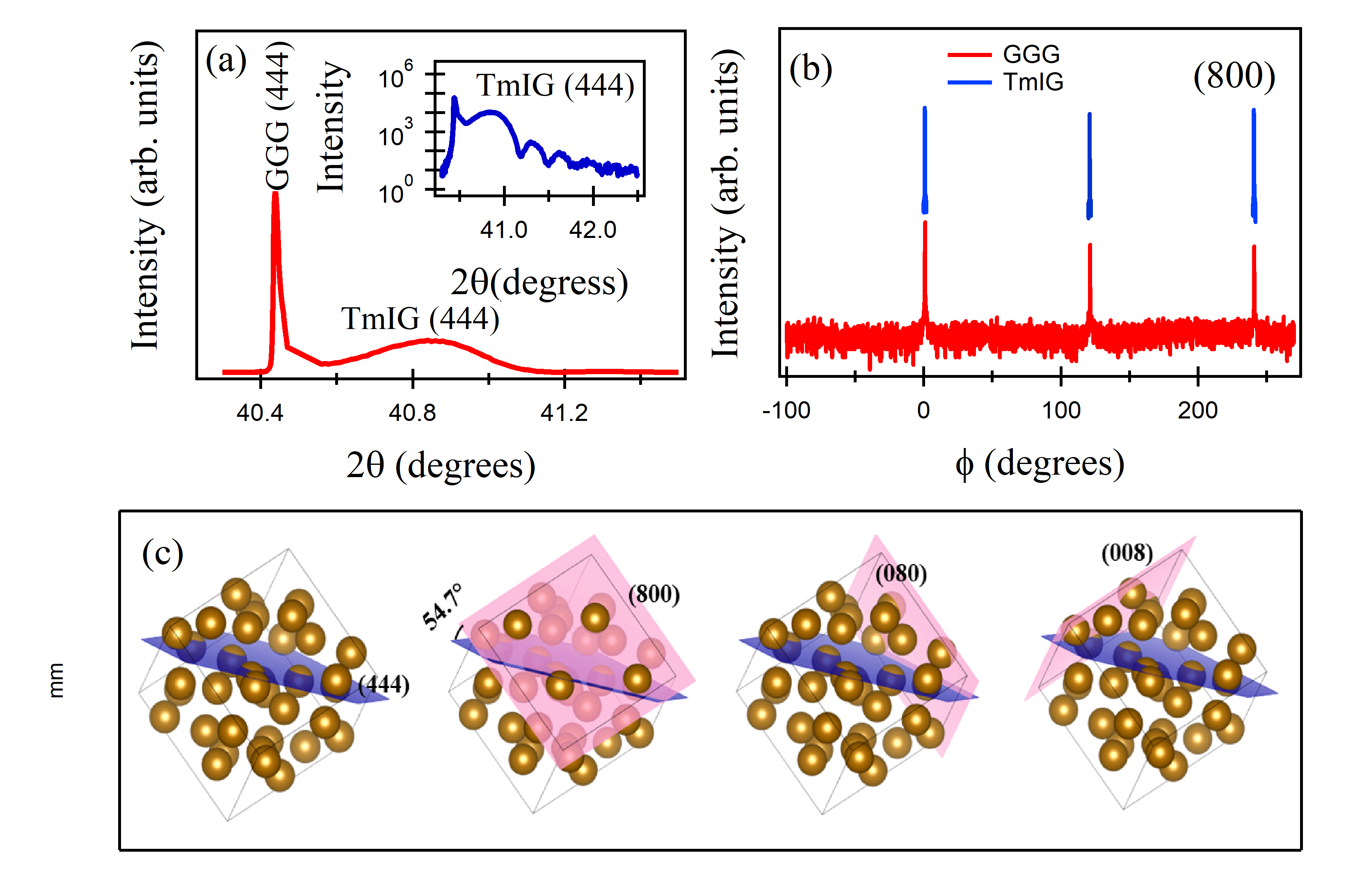}
    \caption{Structural confirmation of the epitaxial TmIG thin film deposited using all-solution-based spin-coating. (a) depicts the $\theta-2\theta$ scan of GIXRD of the substrate and the thin film with the logarithmic inset to present excellent crystallinity, (b) is the $\phi$-scan to confirm the epitaxy with the three-fold symmetry of the (008) plan, and (c) shows the schematic of the three-fold symmetry in (008) plane.}
    \label{fig:X-ray diffraction of the epitaxial TmIG thin film}
\end{figure}

The phase formation of the TmIG thin film deposited using sol-gel-based spin coating is performed using the GIXRD. As the mismatch between the substrate and the thin film is less than a percent, therefore, highly monochromatic 10 keV synchrotron X-rays have been utilized. Figure \ref{fig:X-ray diffraction of the epitaxial TmIG thin film} (a) presents the out-of-plane XRD of TmIG (444) reflection with the substrate GGG (444) highest intensity reflection. Inset Figure \ref{fig:X-ray diffraction of the epitaxial TmIG thin film}(a) shows the logarithmic plot of the intensity to show the excellent interface quality that confirms high crystallinity due to the presence of Laue oscillations \cite{duong2022interfacial}. The interplanar distance of GGG (444) and TmIG (444) are 1.7938 $\pm$ 0.0085 \AA, and 1.7778 $\pm$ 0.0084 \AA, respectively. The lattice constants are 12.4281 $\pm$ 0.0116 \AA, and 12.3145 $\pm$ 0.0116 \AA, for the GGG and TmIG, respectively. Experimental data confirm the smoothness of the interface and the epitaxial growth between the substrate and thin film as shown in Figure \ref{fig:X-ray diffraction of the epitaxial TmIG thin film} (b). The strain because of the mismatch between the two $\Delta\epsilon$ = $\frac{a_{GGG}- a_{TmIG}}{a_{GGG}}$ is 0.88 $\%$ which shows the tensile strain on the layer of TmIG thin film. The tensile strain is the cause of the presence of the PMA in the sample (discussed in further sections). Figure \ref{fig:X-ray diffraction of the epitaxial TmIG thin film} (b) represents the $\phi$-scan of the TmIG thin film. The $\phi$-scan is measured along the (008) Bragg reflection, which is $\psi$ = 54.7$^\circ$ from the (111) Bragg reflection \cite{ma2017room}. The three-fold symmetry in the $\phi$-scan can be observed in Figure \ref{fig:X-ray diffraction of the epitaxial TmIG thin film} (c). The angle difference of 120$^0$ between three-fold symmetry is experimentally observed in $\phi$-scan, which confirms the epitaxy of the deposited thin film \cite{harrington2021back}. The stress ($\sigma$) at the interface is calculated using the following equation \cite{fu2017epitaxial}: 
\begin{equation}
    \sigma = \frac{Y}{1-\nu} \Delta\epsilon
    \label{equ:stress calculation}
\end{equation}

 where, Y is Young's modulus (2.00$\times$10$^{11}$N/m$^2$) and the $\nu$ is Poisson's ratio (0.29) as present in literature \cite{zanjani2020predicting}. The $\sigma$ calculated is 2.573$\times$10$^9$ $\pm$ 2 N/m$^2$.

\begin{figure}
    \centering
    \includegraphics[width=1.0\linewidth]{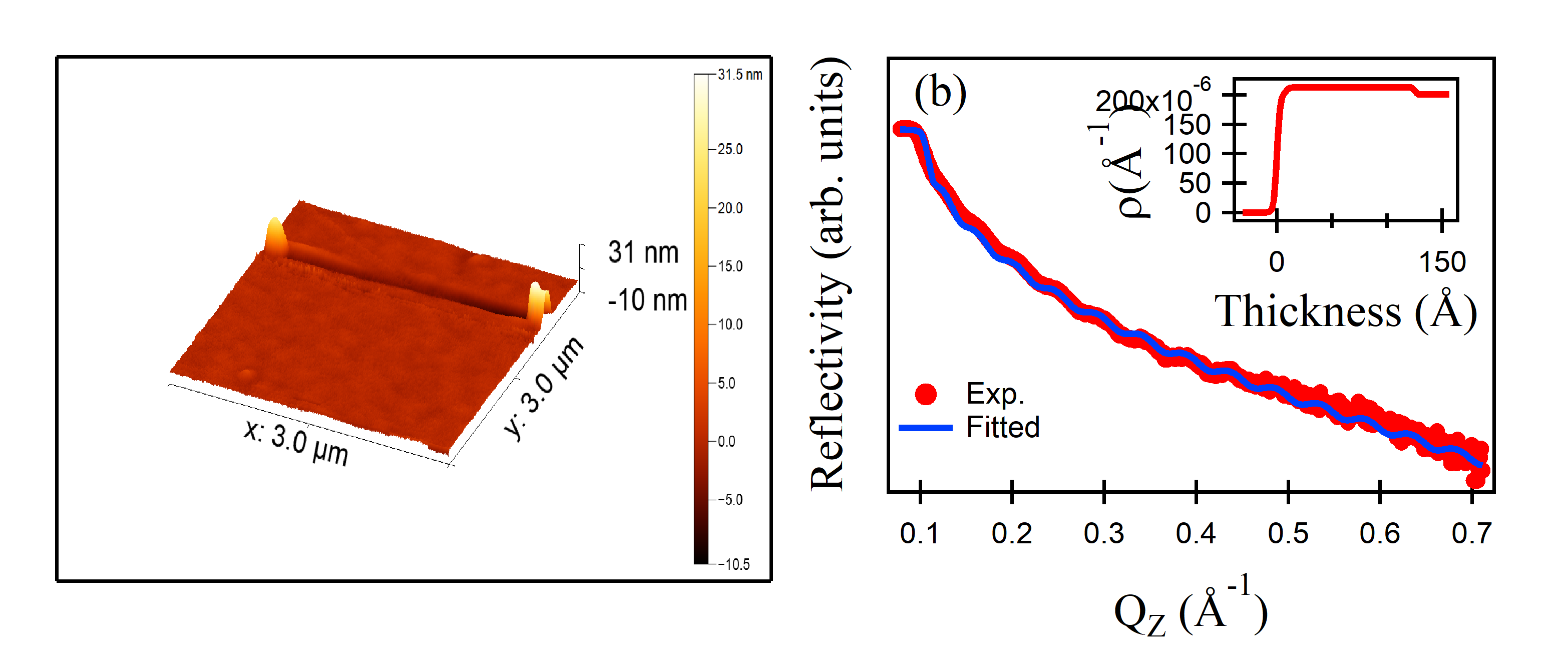}
    \caption{Surface topography and the thickness and roughness study estimation using (a) atomic force microscopy and (b) x-ray reflectivity, respectively.}
    \label{fig:Surface topography of the epitaxial TmIG thin film}
\end{figure}

The substrate film interface quality is essential for the magnonic application. Figure \ref{fig:Surface topography of the epitaxial TmIG thin film} illustrates the topography and structural quality of the deposited thin film using AFM and XRR. Figure \ref{fig:Surface topography of the epitaxial TmIG thin film} (a) illustrates AFM showing smooth topography, estimating mean roughness is $\approx$0.8 nm. Figure \ref{fig:Surface topography of the epitaxial TmIG thin film} (b) illustrates the XRR of the TmIG thin film fitted using Parratt's formalism and that estimated the thickness of the $\approx$12.2 nm, and the interfacial roughness $\approx$0.2 nm, which is excellent. The degree of the crystalline can also be accessed using the presence of Laue oscillations in the inset of Figure \ref{fig:X-ray diffraction of the epitaxial TmIG thin film}(a). The surface roughness estimated is $\approx$0.4 nm, which is in order equivalent to surface roughness estimation using AFM. The topography of the deposited thin film is smooth and suggests homogeneous growth on the substrate. Low surface and interface roughness show the potential of the sol-gel-based spin-coating method to study further application possibilities.

\begin{figure}
    \centering
    \includegraphics[width=1.0\linewidth]{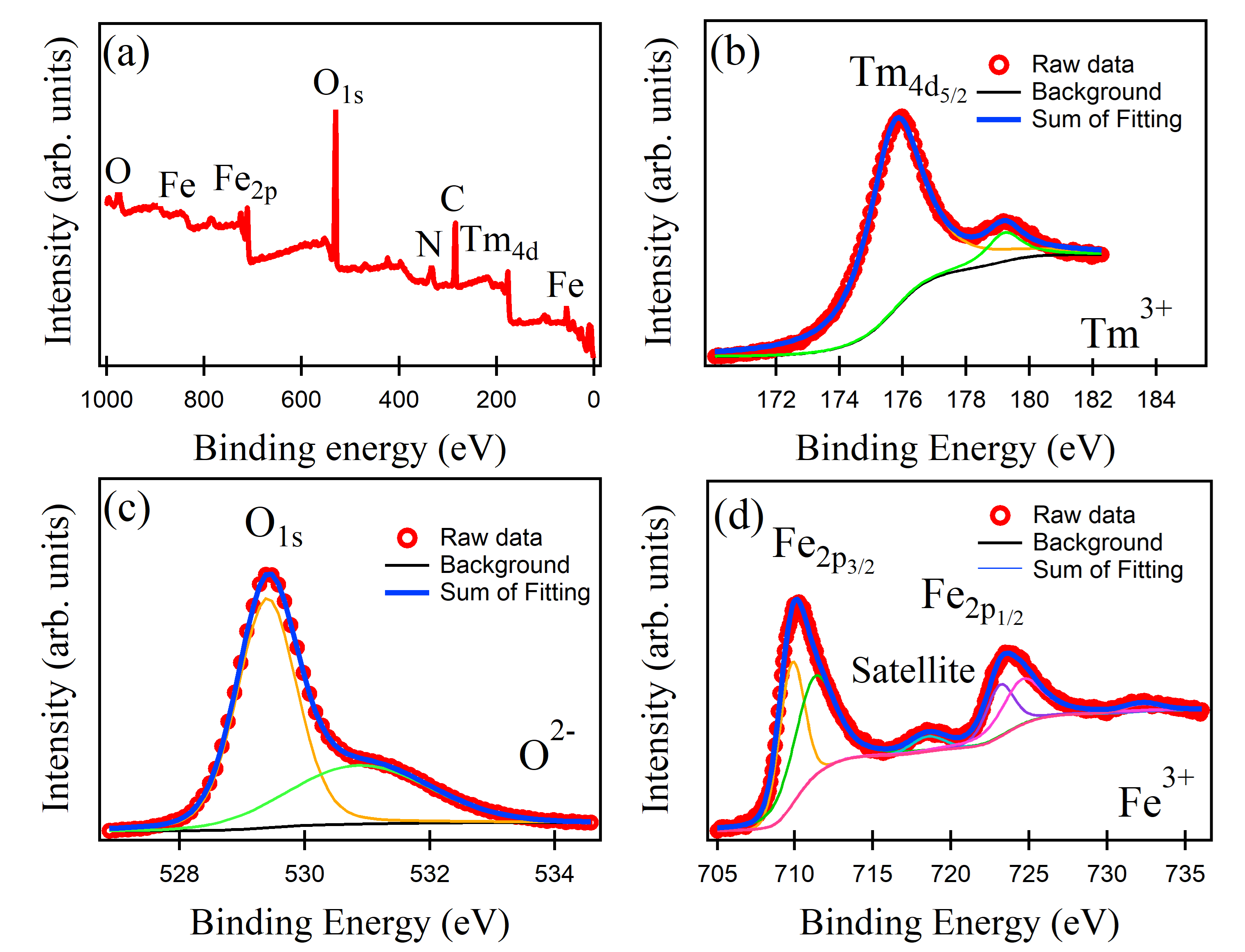}
    \caption{XPS of the TmIG thin film has been presented. (a) survey scan from binding energy 0-1000 eV. High-resolution XPS of the elements (b) Tm$^{3+}$, (C) O$^{2-}$ and (d) Fe$^{3+}$ are fitted and presented with peak component with the Shirley background.}
    \label{fig:TmIG_GGG_XPS}
\end{figure}

The elemental composition is probed using the XPS. Figure \ref{fig:TmIG_GGG_XPS}: depicts the survey scan and the high-resolution XPS of the TmIG thin film. Figure \ref{fig:TmIG_GGG_XPS} (a): plot the survey scan, gives the presence of O, C, Fe, Tm, and N in the thin film. The sample constituted O, Fe, and Tm, but the environmental exposure caused the C and N absorption. The sample is calibrated using the carbon at peak position 284.2 $\pm$ 0.1 eV. Figure \ref{fig:TmIG_GGG_XPS} (b) illustrates the high-resolution spectra of thulium 4d$_{5/2}$ core electrons. Tm$_{4d_{5/2}}$ is observed at 175.7 $\pm$ 0.1 eV which is supported by the literature \cite{chastain1992handbook}. Thulium is present in Tm$^{2+}$ and Tm$^{3+}$ but the most stable valence state is Tm$^{3+}$. The presence of a satellite peak at 179.2 $\pm$ 0.1 is similar to the literature and confirms Tm$^{3+}$ charge state \cite{wang2012band}. Figure \ref{fig:TmIG_GGG_XPS} (c) illustrates the high-resolution spectra of the oxygen 1s core electrons. The main peak at binding energy 529.4 $\pm$ 0.1 eV is because the O$_{1s}$ electrons bind in the TmIG; along with this, the surface contribution of the oxygen is also there at 530.8 $\pm$ 0.1 eV. Figure \ref{fig:TmIG_GGG_XPS} (d) illustrates the high-resolution spectra of the iron 2p core electrons. Iron is present in a 2:3 ratio of octahedral and tetrahedral coordinates in TmIG (space group Ia$\bar3$d). Therefore, the Fe 2p$_{3/2}$ and 2p$_{1/2}$ peak comprise two peaks each. The Fe octahedral (Fe$_{oct}$) peaks 2p$_{3/2}$ at binding energy 709.8 $\pm$ 0.1 eV and 2p$_{1/2}$ at binding energy 723.1 $\pm$ 0.1 eV. The Fe tetrahedral (Fe$_{tetra}$) peaks 2p$_{3/2}$ at binding energy 711.2 $\pm$ 0.1 eV and 2p$_{1/2}$ at binding energy 724.5 $\pm$ 0.1 eV. The theoretical ratio between the area of octahedral and tetrahedral Fe is 2:3. The experimental ratio of the area of peak 2p$_{3/2}$ octahedral and tetrahedral Fe is 0.72, and of peak 2p$_{1/2}$ octahedral and tetrahedral Fe is 0.67 \cite{pan2023angular}. The ratio is very close to the theoretical ratio 2:3, confirming that the quality of the sample is excellent \cite{pan2023angular}. The difference between the core electron peak and the satellite peak is large and equivalent to the 8 eV, which further establishes that Fe is in a 3+ valence state and the stoichiometry is balanced \cite{sharma2023magneticTm}. The atomic percentage of the constituents Tm$^{3+}$, Fe$^{5+}$ and O$^{2-}$ are 16 \%, 26\%, and 58\%, respectively. The atomic percent is calculated using the CASAXPS software \cite{fairley2021systematic}.

\begin{figure}
    \centering
    \includegraphics[width=1.0\linewidth]{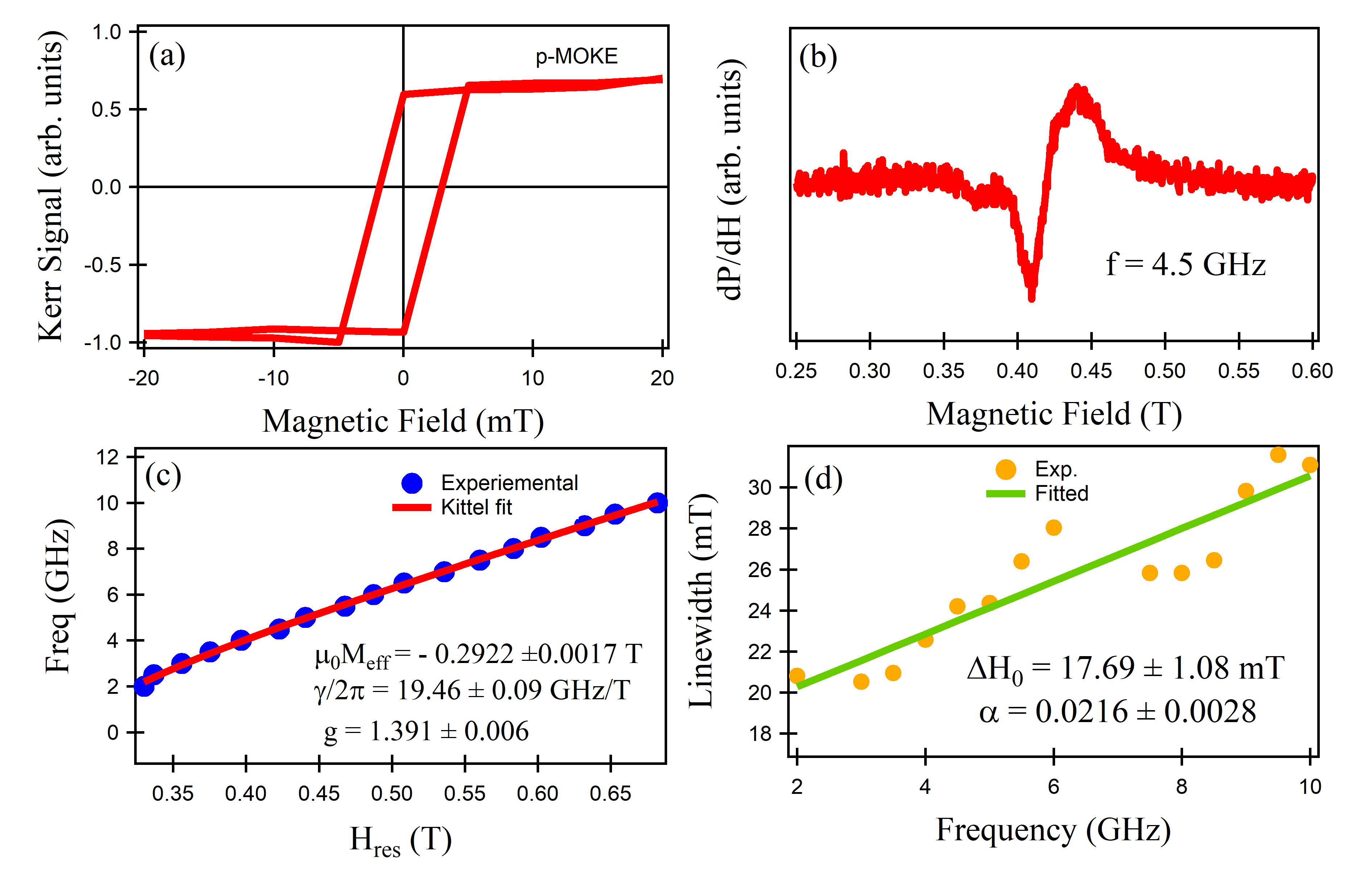}
    \caption{Magnetic Ferromagnetic resonance study of the TmIG thin film (a) Kittel fitting to the resonance magnetic field to study effective magnetic field and gyromagnetic ratio. (b) Linear fit to the linewidth as a function of frequency }
    \label{fig: Magnetic Properties of the epitaxial TmIG thin film}
\end{figure}

As the TmIG has application in magnonics, the magnetic properties of the deposited sample determine its application potential. Figure \ref{fig: Magnetic Properties of the epitaxial TmIG thin film} presents the magnetic behaviour of the deposited all solution-based epitaxial TmIG thin film. Figure \ref{fig: Magnetic Properties of the epitaxial TmIG thin film} (a) illustrates the polar MOKE measurements. The presence of out-of-plane uniaxial anisotropy gives the MOKE signal, which confirms the existence of the perpendicular magnetic anisotropy (PMA) at room temperature in the thin film \cite{quindeau2017tm3fe5o12}. The uniaxial anisotropy (K$_U$) in the thin film is a combination of various components like stress-induced anisotropy (K$_\sigma$), magneto-crystalline anisotropy (K$_{M}$), and shape anisotropy (K$_{S}$) \cite{bhoi2018stress}. Figure \ref{fig: Magnetic Properties of the epitaxial TmIG thin film} (b) illustrates the schematics of the PMA in the film. The stress-induced anisotropy is calculated using the formula as follows:

\begin{equation}
    K_\sigma = -\frac{3}{2} \lambda_{111} \sigma
    \label{equ:stress-induced anisotropy}
\end{equation}
where, $\lambda_{111}$ is the magnetostriction constant (-5.2 $\times$10$^{-6}$) of the TmIG as present in literature \cite{zanjani2020predicting}. The value of the estimated stress-induced anisotropy (K$_\sigma$) is 20.07$\times$10$^3 \pm $ 5.71 $10^{-2}$ N/m$^2$. Shape anisotropy (K$_{S}$) estimated is 0.49 $\times 10^{3}$ N/m$^2$. The cubic anisotropy constant (K$_1$) value is taken from the literature\cite{quindeau2017tm3fe5o12} is -1.1$\times 10^{3}$ N/m$^2$. The final uniaxial anisotropy value is estimated as follows:

\begin{equation}
    K_U = -\frac{K_1}{12} + K_\sigma + K_{S}
\end{equation}

where, K$_1$/12 is K$_{M}$. The estimated K$_U$ from the strain is 20.11 $\times 10^{3}$ N/m$^2$.

The magnetic study of the TmIG is also performed using FMR. It probes the precession of the moments along the external field, and this precession resonates with the applied frequency. The absorption of that frequency at a particular magnetic field gives the resonance magnetic field and the linewidth of the absorption, which signifies the moment's precession and energy dissipation, respectively. Figure \ref{fig: Magnetic Properties of the epitaxial TmIG thin film}: depicts the FMR results, (a) plots in-plane the resonance magnetic field (H$_{res}$) as a function of frequency and fitted using Kittel equation, (b) illustrates the linewidth as a function of the frequency. Kittel equation \cite{chang2017role,zhang2022strong} is as presented below: 

\begin{equation}
	\label{Kittel_equation}
		f = \left(\frac{\gamma}{2\pi} \right) \sqrt{H(H+\mu_0M_{eff})}
\end{equation}	

The effective magnetization ($\mu _0$M$_{eff}$) estimated is -0.292 $\pm$ 0.003 T. As the PMA is confirmed with the polar MOKE, the negative value of $\mu _0$M$_{eff}$ shows that anisotropy dominates the saturation magnetization. $\mu _0$M$_{eff}$ is composed of saturation magnetization ($\mu _0$M$_{S})$ and the anisotropy field (H$_U$) in following equation\cite{bhoi2018stress}:

\begin{equation}
   \mu _0M_{eff} = \mu _0M_{S} - H_U 
   \label{equ:Meffective}
\end{equation}
 In literature, the value of saturation magnetization ($\mu _0$M$_S$) of bulk TmIG is 0.1244 T \cite{wu2018high}. The anisotropy Field ($H_U$) is estimated to be 0.4167 T. The anisotropy constant (K$_U$) is calculated by substituting the anisotropy field and saturation magnetization as present in the following equation:
 \begin{equation}
     K_U = \frac{H_U \times M_S}{2}
     \label{equ:Anisotropy constant as HU and MS}
 \end{equation}

 K$_U$ is estimated by substituting, the values is 20.6$\times$10$^3$ $\pm$ 0.2$\times$10$^3$ N/m$^2$. K$_U$ estimation by FMR is equivalent to the K$_U$ calculated with the strain in GIXRD. This value is higher than the literature due to the self-growth of the TmIG/GGG and form a better interface. The value of the gyromagnetic ratio is 19.46 $\pm$ 0.09 GHz/T, which is lower than the free electron because of the high spin-orbit coupling of the thulium ions. The Land\'{e} g-factor estimated is 1.391 $\pm$ 0.006 from the gyromagnetic ratio. The Land\'{e} g-factor is smaller than the free electron as well as the value present for TmIG in literature\cite{zhang2022strong}. The presence of low Land\'{e} g-factor can be due to the presence of high anisotropy present in TmIG/GGG thin film.
 The uniform mode is generated while the moment precession relaxes by the dissipating energy due to extrinsic and intrinsic factors. Extrinsic factors are like defects and electro-electron interaction, and intrinsic factors are like two-magnon interaction and high spin-orbit interaction. These factors can be obtained from the linear fitting of the magnetic linewidth as a function of the applied frequency. Figure \ref{fig: Magnetic Properties of the epitaxial TmIG thin film} (d) depicts the linewidth ($\Delta H$) as a function of the applied frequency. Yellow dots are obtained by analyzing the experimental FMR data as a function of the applied field (the intensity of the dP/dH is low, which causes the deviation), and the green line fits them linearly. The relation of linewidth, intrinsic, and extrinsic damping parameters is as follows:

 \begin{equation}
     \Delta H = \Delta H_0 + \frac{4\pi \alpha}{\gamma} f
 \end{equation}
 
 The $\Delta H_0$ is the extrinsic part of the TmIG energy dissipation, and the fitted value is the $\Delta H_0$ is 17.69 $\pm$ 1.08 mT. The intrinsic dissipation is stated as the Gilbert damping parameter ($\alpha$) = 0.0216 $\pm$ 0.0028. Gilbert damping parameter is of the same order as the present in literature in which samples are prepared with sophisticated methods \cite{quindeau2017tm3fe5o12}.  

\begin{table}[h!]
  \begin{center}
  	\caption{The comparison of the present data with the literature.}
\label{tab:tablecomparision}
\resizebox{1.0\columnwidth}{!}{
\begin{tabular}{c c c c c c}
\hline

     Substrate & Deposition Method & Thickness & Damping Parameter & Anisotropy constant  & reference \\
      &  & (nm) & & (10$^3$ N/m$^2$) & \\ 
     \hline
     GGG (111) & Sol-gel-based Spin-coating & 12.2 & 0.0216 & 20.6 & present\\
     GGG (111) & PLD & 29 & 0.013 & 9 & \cite{rosenberg2021magnetic}\\
     GGG (111)  & PLD & 20 - 300 & 0.02 & 10 & \cite{ciubotariu2019strain} \\
     GGG (111) & sputtering & 9 & 0.013 & - & \cite{wu2018high}\\
     GGG (111) & PLD & 8 & - & 11.88 & \cite{quindeau2017tm3fe5o12}\\
     \hline
\end{tabular} }
\end{center}
\end{table}

This article presents a cost-effective method and to understand it the literature is compared with experimental observation in table \ref{tab:tablecomparision}. All the samples are grown on the GGG (111) substrate, and the thicknesses are different, but the Gilbert damping parameter is of the same order. The estimated K$_U$ is higher compared to the literature and is supported by the lower value of the Land\'{e} g-factor of the present work.

In conclusion, sol-gel-based spin coating is utilized to deposit the epitaxial thulium iron garnet (TmIG) thin film on the GGG substrate. The elemental analysis confirms the stoichiometric deposition with the low interface and the surface roughness. The presence of the perpendicular magnetic anisotropy of this all-solution method deposited TmIG due to the stress-induced anisotropy. Due to the presence of the high spin-orbit coupling gives rise to the lower gyromagnetic ratio and Land\'{e} g-factor, which is well matched with the literature. The intrinsic and extrinsic dissipation factors of TmIG present a potential for the deposition method. With further improvements, this cost-effective method of deposition has the potential to be used for magnonics applications.

\section*{CRediT authorship contribution statement}
\textbf{Rajnandini Sharma}: Conceptualization (lead); Synthesis (equal); Data curation (lead); Formal analysis (lead); Project administration (supporting); Validation (equal); Visualization (lead); Writing – original draft (lead); Writing – review \& editing (supporting).\textbf{Pawan Kumar Ojha}: Synthesis (equal); Formal analysis (supporting). \textbf{Simran Sahoo}: Formal analysis (supporting); Writing – review \& editing (supporting).\textbf{Rijul Roychowdhury}: Experimental contribution (GIXRD). \textbf{Shrawan K. Mishra}:  Funding acquisition (lead); Project administration (lead); Supervision (lead); Validation (equal); Visualization (supporting); Writing – original draft (supporting); Writing – review \& editing (lead).

\section*{Data Availability Statement}
The data that support the findings of this study are available from the corresponding author upon reasonable request.

\section*{Declaration of Competing Interest}
The authors declare that they have no known competing financial interests or personal relationships that could have appeared to influence the work reported in this paper.

\section*{Acknowledgement}
This work is financially supported by the Nano Mission program, DST, India project No. IIT(BHU)/R$\&$D/SMST/18-19/09. RS acknowledges the DST INSPIRE for the INSPIRE fellowship. The authors are thankful to the Saha Institute of Nuclear Physics, Kolkata, for facilitating the experiments at the GIXS Beamline (BL-13), Indus-2, RRCAT, Indore, and for the technical support received during the beamtime. The authors are thankful to Dr. V. R. Reddy and UGC-DAE CSR, Indore, for the MOKE experiment.

\bibliography{references} 

\end{document}